# Global DIC-based sample-detector geometry refinement for accurate EBSD indexing

Claire Griesbach[1], Dennis M. Kochmann[1]

[1] Mechanics & Materials Laboratory, Department of Mechanical and Process Engineering, ETH Zürich, 8092, Zürich, Switzerland

**Abstract**

Electron backscatter diffraction is a powerful tool for mapping crystallographic microstructures. However, the primary crux to improving orientation accuracy and applying the technique to challenging materials lies in the correct calibration of the sample-detector geometry. Many approaches have aimed at overcoming this barrier through various pattern center calibration strategies, but the pattern center only defines part of the sample-detector geometry. Here, we present a DIC-based geometry refinement method that obtains a single map-consistent sample-detector geometry, refining both the pattern center and sample/detector angles. We effectively decouple the local orientation changes from the global geometry effects on the Kikuchi patterns by calculating the consistent map-wide simulated-to-experimental pattern shifts associated with global geometry parameter errors. Using single-crystal silicon and barium titanate (a material possessing six pseudosymmetric variants) as model materials, we demonstrate improved map-wide orientation consistency and more robust discrimination of pseudosymmetric variants than the Nelder-Mead and Differential Evolution optimization strategies.

## 1. Introduction

Electron backscatter diffraction (EBSD) is a powerful microstructural characterization technique that provides spatial maps of a sample's crystal structure. Electron diffraction by the crystal lattice is captured on an EBSD detector, forming a Kikuchi pattern. Automatic Hough indexing of Kikuchi patterns (to determine the crystal orientation) was established more than 30 years ago [1] and is still widely used today. Recent advancements in EBSD analysis have enabled the quantification of lattice strain through high resolution EBSD (HR-EBSD) [2,3], spatial resolutions of a few nanometers through Transmission Kikuchi Diffraction (TKD) [4], and indexing of highly deformed crystals (exhibiting poor pattern quality) and pseudosymmetric materials using simulated pattern-matching approaches [5–9].

While these advanced indexing techniques show great improvements in the angular precision (the relative orientation between neighboring patterns), the accuracy of the absolute crystal orientation remains limited. Hough indexing can typically achieve an angular precision of ~0.5° and an absolute orientation accuracy of 1-2° [10,11]. HR-EBSD has been shown to achieve angular precisions down to ~0.01° [3,12]. A few recently proposed approaches perform global DIC on Kikuchi patterns to estimate the relative strain down to $10^{-4}$ or ~0.006° [13–16]. Although these approaches enable high angular precision, the absolute orientation accuracy is limited by accurate calibration of the sample-detector geometry. An analysis of the orientation error obtained by the dictionary indexing technique reported the potential for a 0.2° orientation accuracy, with the caveat that the error in detector geometry can be reduced to 0.1% in the projection center and 0.1° in the sample tilt [17]. When simulated patterns are used for HR-EBSD, a typical 0.5% error in the pattern center produces phantom strains on the order of $5 \times 10^{-3}$, which is large in comparison to expected lattice strains [12].

Pattern-matching approaches also provide the potential to characterize microstructures of materials with crystal structures possessing pseudosymmetries. Kikuchi patterns of pseudosymmetric (PS) variants (described by a rotation of the unit cell) contain minute differences in the intensity distributions, which are typically due to dynamic diffraction and not from geometrically-defined Bragg diffraction. This means that Hough-based EBSD indexing cannot distinguish between PS variants, and patterns simulated with dynamical diffraction must be used for indexing [18,19]. Precise calibration of the pattern center is crucial for accurate discrimination between PS variants using pattern-matching approaches [9]. As the technique progresses to analyze more challenging materials, precise calibration of the sample-detector geometry parameters becomes increasingly crucial to obtain accurate indexing results.

The Kikuchi pattern that is captured by the EBSD detector is dependent on both the local crystallography at the point on the sample where the electron beam is focused (interaction point) and on the geometry between the sample and the detector. This sample-detector geometry defines a global reference frame by which the crystal orientations can be defined. As such, accurate determination of the crystal orientations from the Kikuchi patterns relies on knowledge of the precise geometry. Typically, six parameters define the geometry: three parameters characterizing the point on the detector positioned closest to the interaction point on the sample (i.e., the pattern center) and three angular parameters (the sample tilt, detector tilt, and detector azimuthal angle).

Pattern center calibration has received much attention, and there are several available methods that can be grouped into two main categories: (i) pre-scan calibrants, which are performed system-wide prior to data acquisition, and (ii) post-scan calibrants, which are performed after data acquisition using the stored Kikuchi patterns. The technique standard to most commercial EBSD setups is a pre-scan technique, which iteratively alters the projection geometry to best fit the expected interplanar angles for the crystal as measured from the Hough-identified bands [20,21]. Often, the moving-screen approach [22] is performed on an easily indexable single crystal at various detector positions to produce a system-wide calibration that can be read from for an initial calibration setting. These common techniques are reported to achieve an accuracy in pattern center of 0.5% [12,20]. Other pre-scan calibrants include the shadow-casting [23] and single crystal reference [24] techniques, which require special equipment or samples positioned with precisely the same geometry as the sample to be analyzed.

Alternatively, post-scan calibration techniques are computationally performed after data acquisition, provided the original patterns were saved. The virtual screen technique relies on the fact that a pattern projected onto the Kikuchi sphere with a properly tuned pattern center will have bands with parallel edges aligned with great circles on the sphere [25]. Given the advancements in Kikuchi pattern simulations, several computational optimization techniques have been used to refine the pattern center and orientation until a maximum correlation between the experimental and simulated patterns is obtained [26–29]. However, small changes in pattern center values have been shown to effectively compensate for changes in orientation, resulting in a sloppy optimization landscape [26,30]. Global optimization algorithms, like the Differential Evolution algorithm, perform better than local optimization algorithms like Nelder-Mead, which can fail to converge or converge incorrectly in the sloppy optimization landscape [26,27]. With any numerical optimization strategy, convergence to the best pattern center for the map is typically accomplished by taking the average pattern center over many independently optimized patterns in the map, resulting in pattern center accuracies of down to 0.0002–0.008% of the detector width [26]. However, this averaging approach is not suitable for large maps where the pattern center varies with map geometry [28] or where close PS variants cannot be distinguished and the optimized pattern center incorrectly varies between variant orientations [31]. Additionally, because these approaches optimize patterns independently, they do not explicitly enforce a physically consistent geometry for the entire map.

While extensive work has been devoted to pattern center calibration, there are up to four additional sample-detector geometry parameters that are usually not rigorously calibrated: the detector tilt, sample tilt, detector azimuthal angle, and detector twist. These angles are typically considered to be fixed and known from the EBSD detector installation or SEM stage positioning readout. However, it has been shown that errors in these angular parameters result in comparable errors in the absolute orientations (e.g., a 2° error in the detector tilt causing a 2° disorientation) [17,27,32].

In this paper, we propose a method to simultaneously refine all sample-detector geometry parameters, including the detector and sample angles. Taking inspiration from the global DIC approach to HR-EBSD strain analysis [13–16], we use local DIC to quantify the displacement fields between simulated and experimental patterns. In contrast to HR-EBSD methods, which are fundamentally aimed at recovering relative deformation or disorientation, we use these local displacement measurements to identify the consistent map-wide pattern shifts associated with errors in the global sample–detector geometry. Exploiting these spatially coherent signatures enables the effects of global geometry to be decoupled from local orientation changes. Two model materials, single-crystal silicon and single-crystal barium titanate (which possesses six pseudosymmetric variants), are used to demonstrate the effectiveness and reliability of the method in comparison with state-of-the-art pattern-center optimization techniques.

## 2. Methods

### 2.1. Global vs local pattern shifts and the coupling between orientation and geometry

In this section, we demonstrate the need for an approach which effectively decouples local crystallographic changes from the sample-detector geometry. Using *Kikuchipy*, we simulate a Kikuchi pattern from a barium titanate (BTO) master pattern, using an initial known detector geometry and crystal orientation. Then, we perturb one of the six geometry parameters, the crystal orientation, or apply a pseudosymmetry operation. *Kikuchipy*'s *refine_orientation_projection_center* function is used with a Nelder-Mead optimization algorithm to attempt to find a combination of orientation and pattern center values which bring the simulated pattern with the initial perturbed parameters in alignment with the original pattern.

Figure 1 shows the results of that approach. The left side of the figure displays the perturbed parameter (and amount) and plots showing the local intensity differences between the original and perturbed pattern. Towards the right, we show the changes in pattern center and orientation optimized through the refinement algorithm and the intensity difference plots after refinement. The normalized cross-correlation (NCC) values are also displayed on the intensity difference plots. The intensity difference plots and NCC values after refinement show nearly perfect pattern alignment for all parameters. The refinement of orientation and pattern center perturbation cases show almost perfect recovery of the true pattern center and orientation. This shows that the optimization algorithm performs well with these perfect simulated patterns, despite the sloppy optimization landscape from the six coupled orientation and pattern center components.

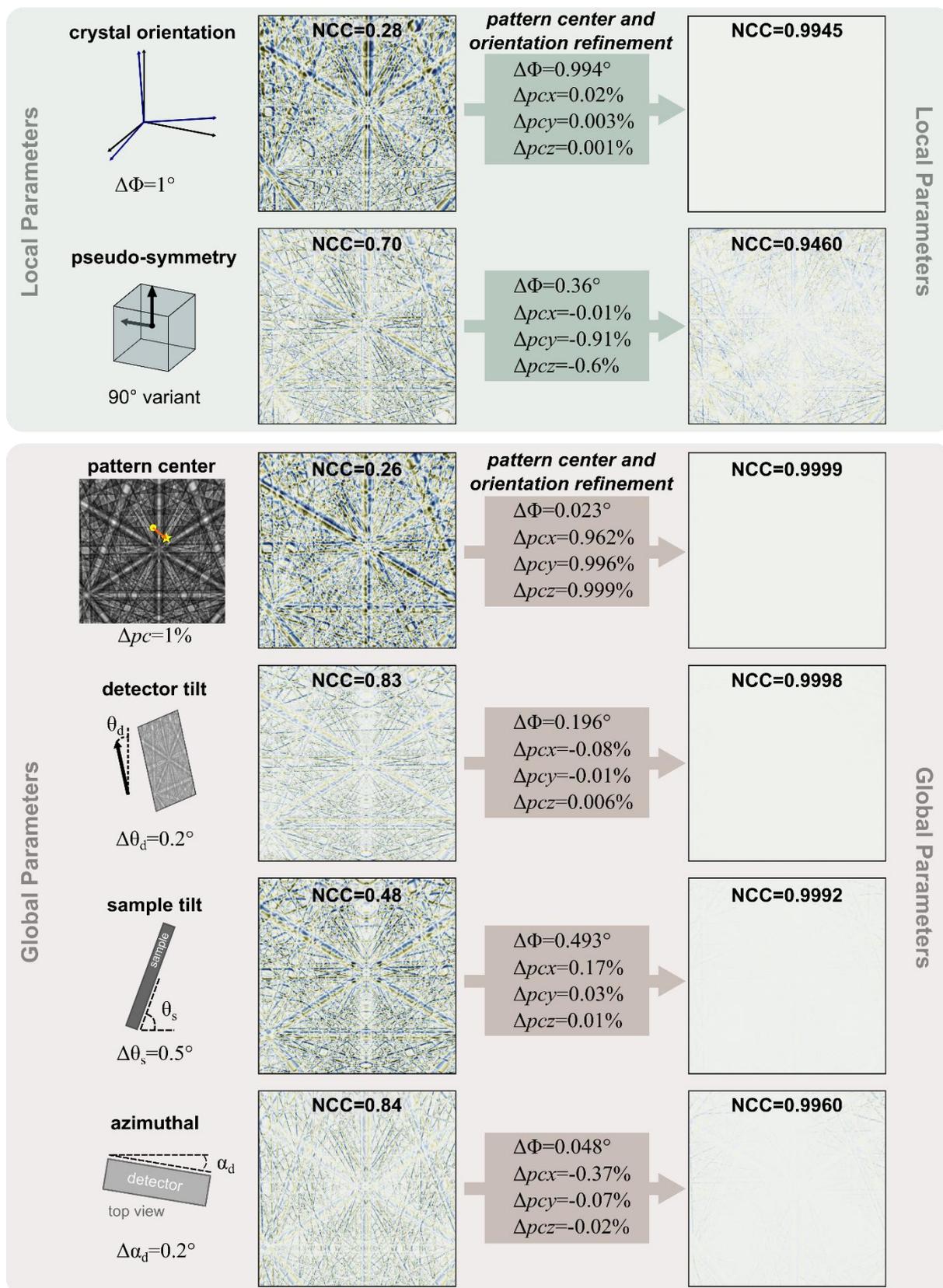

**Figure 1: Coupling between pattern shifts due to local orientation and global geometry changes**

The artificial success of the pattern center and orientation refinement in the other parameter cases is more concerning. For example, the pseudo-symmetry case (Figure 1 – second row) shows that an incorrect refinement of geometry and orientation can yield incorrect identification of a PS variant type, with an NCC score of 0.95 when the 90° PS variant pattern should have an NCC score of 0.7. Furthermore, we show that any errors in detector angles or sample tilt can be incorrectly compensated for with a crystallographic orientation change or pattern center. Refinement of these patterns results in almost perfect alignment with NCC scores greater than 0.99 (Figure 1 – "Global Parameters").

This analysis demonstrates that pattern-wise optimization of the pattern center and orientation severely limits the accuracy of EBSD orientation determination. For most cases, orientation errors on the order of a degree are inconsequential for microstructural characterization. However, if precise orientations or discernment between PS variants are required, decoupling between the orientation solution and the sample-detector geometry is required. The method we propose in the following sections effectively decouples the orientation and geometry by considering the consistent global shifts in Kikuchi patterns across the EBSD map when refining the geometry.

*2.2. Using average displacement fields to characterize geometry-related pattern misalignments*

To isolate the effects of the global geometry from the pattern-wise orientation changes, we analyze many patterns together across the EBSD map. Since shifts between individual experimental and simulated patterns could be due to errors in geometry or the indexed orientation, calculating an average shift (or displacement field) over several different patterns highlights the global geometry-related misalignment and negates the effects of local orientation errors. Points distributed across the map are selected (either using an equally spaced grid or manually), ensuring broad representation from different crystal orientations and regions on the sample. A narrow selection from one crystal orientation or region on the map does not effectively decouple the global geometry from the orientation. With this selection of $N$ points ( $i = 1:N$ ), the initial geometry and orientations are used to simulate patterns.

An example pair of experimental and simulated patterns is shown in **Figure 2**A. We adopt the digital image correlation (DIC) method to quantify the observed pattern shift. For each pattern pair (simulated pattern $\mathbf{A}_i$ and experimental pattern $\mathbf{B}_i$), we find the displacement field which brings the patterns into optimal alignment. To accomplish this, both patterns are subdivided into a square grid of regions of interest (ROIs). (The patterns were previously cropped to a square fully filled with the pattern.) The number of ROIs ($R$) is set depending on the performance of the image registration algorithm (i.e., making sure that multiple features can be detected within each ROI). Feature detection is performed using *OpenCV*'s *goodFeaturesToTrack* algorithm, which finds the most prominent corners in the image [33]. Within Kikuchi patterns, these "corners" primarily correspond to the intersection points between bands, as shown in Figure 2A. To improve registration of the same points in the experimental and simulated patterns, image processing is applied prior to feature detection. Pixel binning is useful to reduce intensity variations within the bands. For pattern-matching indexing, this detailed intensity information is necessary, but for feature recognition the intersection points between bands can be more robustly tracked when local intensity differences are averaged out. Since the experimental pattern is typically noisier and less detailed than the simulated image, Gaussian noise is applied to the simulated pattern.

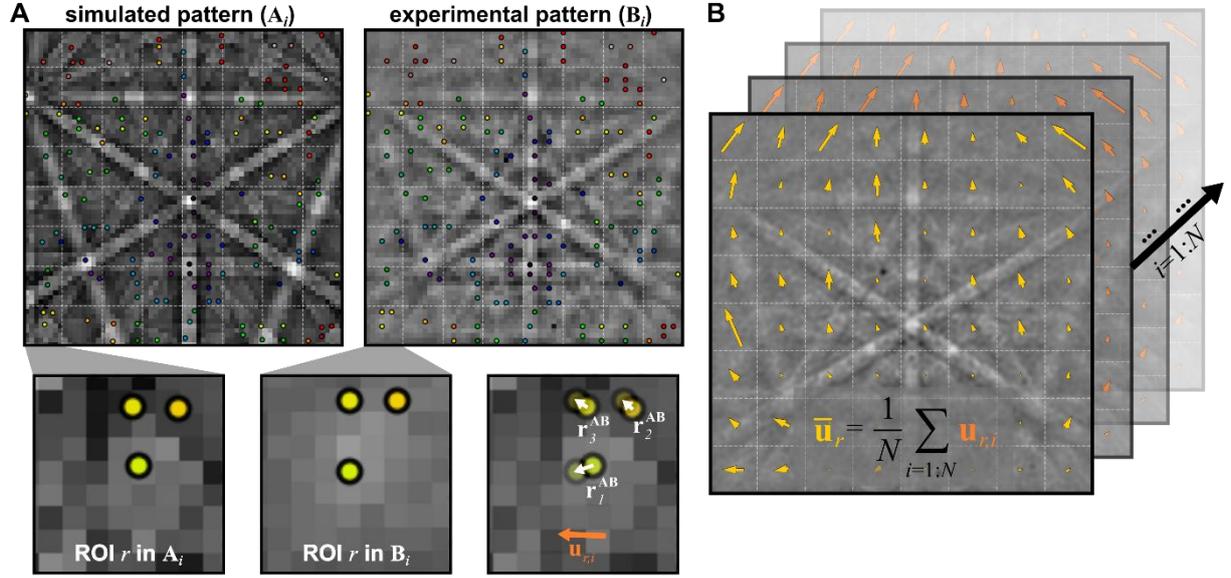

**Figure 2: DIC-based calculation of the displacement field representing pattern misalignment:** (A) Example pair of simulated and experimental patterns (with pixel binning of 4), divided into a grid of ROIs (white dashed lines) and feature pairs identified (same colored dots between the images); an example ROI is magnified from each pattern below, with the third panel showing the displacement vectors calculated from the feature pairs and the average displacement vector for this ROI shown in orange; (B) a representative stack of patterns with calculated per-pattern displacement fields (orange), which are used to calculate the average displacement field (yellow), quantifying the consistent misalignment from global geometry parameter errors.

Once feature points have been detected in the experimental and simulated images, the point pairs between the images are identified using *OpenCV*'s *calcOpticalFlowPyrLK* algorithm, which uses the Lucas-Kanade method to identify the pattern of apparent motion of image objects between two consecutive frames (optical flow) [34]. An example of the feature pairs identified between the simulated and experimental images is shown in Figure 2A. To filter out any erroneous pairings, a pair is not used for the displacement field calculation if the feature pairings cross more than a threshold number of ROIs (typically two). For each remaining pair, the vector between the feature point on the simulated image to the corresponding point on the experimental image is calculated ($\mathbf{r}_j^{AB}$). Then, for each ROI, all vectors which have at least one of their endpoints within the ROI are averaged to yield the displacement vector for that ROI:

$$\mathbf{u}_{r,i} = \frac{1}{c_{r,i}} \sum_{j \in r} \mathbf{r}_j^{AB} \tag{1}$$

for all $j$ within ROI $r$ and image $i$. The number of vectors ($c_{r,i}$) used in the averaging per ROI are also recorded, since this is later used to weight the calculated displacement vector as a measure of accuracy

(a single vector may be erroneous). Finally, the average displacement vector for each ROI is computed by averaging the ROI displacement vectors across images (Figure 2B):

$$\bar{\mathbf{u}}_r = \frac{1}{N} \sum_{i=1:N} \mathbf{u}_{r,i} . \tag{2}$$

Figure 2B shows the average displacement field plotted on a representative Kikuchi pattern. The displacement vectors are not randomly oriented: instead, a pattern emerges with most vectors in the center of the pattern pointing upwards, while those towards the top corners also have an inward facing horizontal component and those towards the bottom corners also have an outward facing horizontal component to the displacement vectors. This coherent pattern in the average displacement field reveals that there is a consistent misalignment for all patterns in the map, which can be attributed to errors in the global sample-detector geometry parameters. This global misalignment is quantified through a residual field vector **b**, which is the stack of average displacement vectors (calculated from Eq. (2)) for all ROIs:

$$\mathbf{b} = [\bar{\mathbf{u}}_1^T, \bar{\mathbf{u}}_2^T, ..., \bar{\mathbf{u}}_R^T] . \tag{3}$$

Now that the effects of the geometry errors have been identified, we must find the combination of geometry parameter changes which correct for these global misalignments.

*2.3. Geometry parameter sensitivities*

Let us quantify the effects of changing each of the six geometry parameters on the Kikuchi pattern. Figure 3A shows a schematic of the sample-detector geometry, with the six individual geometry parameters identified. Three parameters (*pcx*, *pcy*, and *pcz*) define the pattern center (i.e., the point on the detector positioned closest to the interaction point on the sample). Three angular parameters—$\theta_d$, $\alpha_d$, and $\theta_s$—define the detector tilt, detector azimuthal angle, and sample tilt, respectively. Although the angles are typically considered to be fixed and known from the EBSD detector installation or SEM stage positioning readout, they may not be known sufficiently accurately in practice. We will show that small deviations in these parameters can affect the simulated pattern projection in ways that cannot be replicated by shifts in the pattern center or crystal orientation. Therefore, they are included in the geometry refinement method but can be easily kept constant in the code, when desired. We define a parameter vector $\mathbf{p} = [pcx, pcy, pcz, \theta_d, \alpha_d, \theta_s]$, which includes all geometry parameters used in the refinement. The current geometry is given by $\mathbf{p}_0$. For each geometry parameter $p_k$ (where $k=\{1,...,6\}$), we perturb the current geometry by a step $\delta_k$ in each direction—creating two new geometries per parameter:

$$\mathbf{p}_{k+} = \mathbf{p}_0 + \delta_k \mathbf{e}_k \quad \text{and} \quad \mathbf{p}_{k-} = \mathbf{p}_0 - \delta_k \mathbf{e}_k . \tag{4}$$

Patterns are simulated with these two new geometries and compared to the corresponding experimental patterns to construct residual vectors from the average displacement fields. These residual field vectors ($\mathbf{b}_{k+}$ and $\mathbf{b}_{k-}$) describe the resultant pattern shifts from perturbing a single geometry parameter up or down. Next, the Jacobian matrix ($\mathbf{J} = \frac{\partial \mathbf{b}}{\partial \mathbf{p}}$) is computed, representing the linear sensitivity of perturbing each geometry parameter. The columns of the Jacobian sensitivity matrix $\mathbf{J}_{:,k}$ are computed from the residual vectors using a central-difference derivative approximation:

$$\mathbf{J}_{:,k} = \frac{\mathbf{b}_{k+} - \mathbf{b}_{k-}}{2\delta_k} . \tag{5}$$

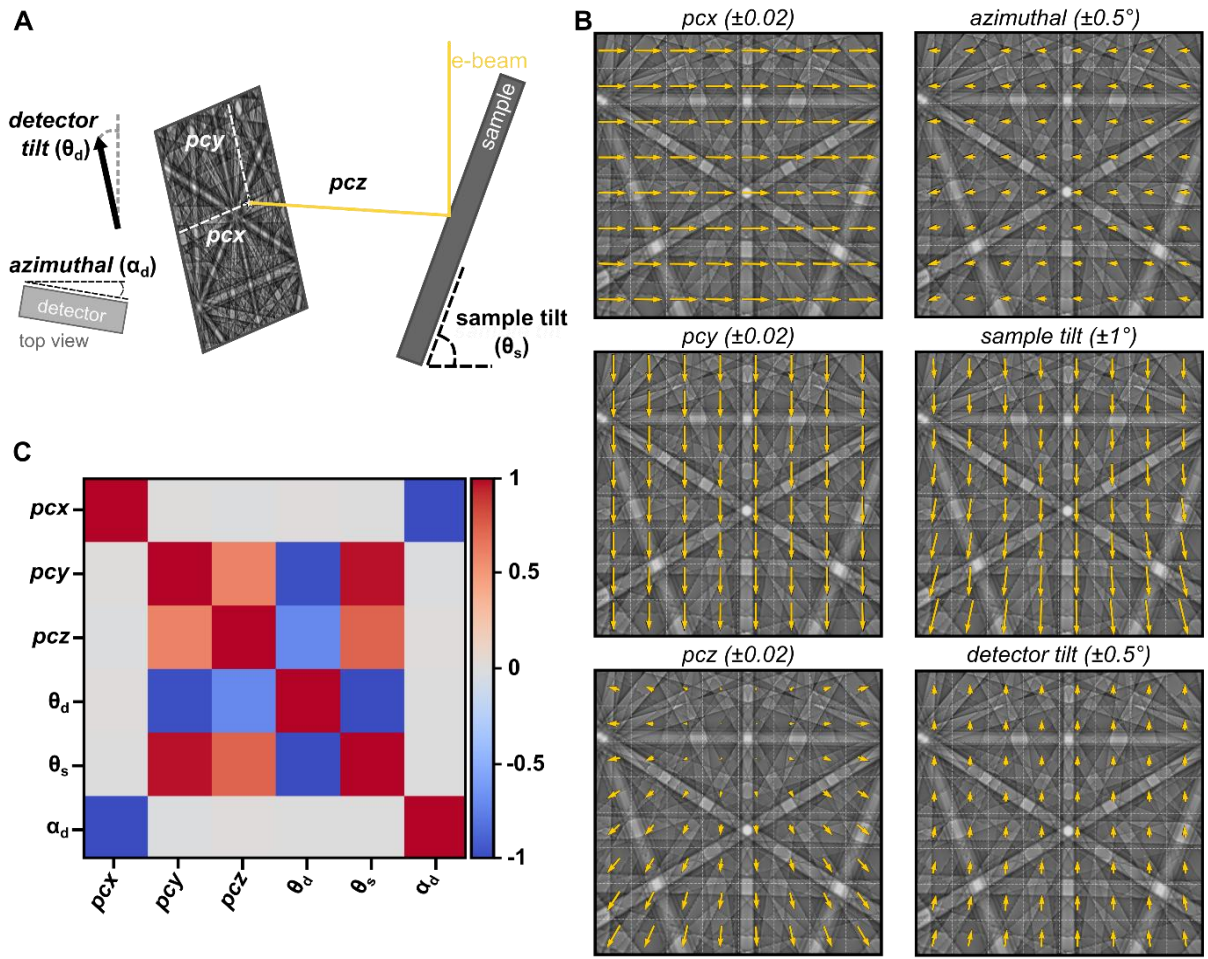

**Figure 3: Sample-detector geometry parameter sensitivities:** (A) schematic showing sample-detector geometry parameters; (B) geometry parameter sensitivities obtained from perturbing one parameter at a time by a specified step (displacement field arrows scaled up for visibility); (C) coupling matrix with values from -1 to 1 showing the degree of coupling between each parameter pairing.

Visual examination of the vector fields obtained from perturbing each parameter individually (Figure 3B) reveals that each parameter has a unique sensitivity, while there is some coupling between parameters. To quantify any coupling, we compute the scalar product of sensitivity vectors for each parameter pair combination ($\mathbf{J}_{:,k} \cdot \mathbf{J}_{:,l}$). This coupling matrix is plotted in Figure 3C. From the coupling matrix and vector fields for the sample and detector tilt perturbations, we see that these two geometry parameters have a strong inverse relationship. Both parameters are also coupled with the *pcy* parameter (and, to a lesser degree, *pcz*), since all these geometry parameters have strong vector fields in the *y*-direction. The detector azimuthal angle is strongly inversely coupled with the *pcx* component of the pattern center, since changing either parameter primarily results in an *x*-directional shift of the pattern. Although the angular parameters all exhibit coupling with the pattern center components, there are also distinct field components that cannot be replicated by changing only the pattern center. For example, the sample and detector tilt fields have localized dilation (or contraction) in the horizontal direction near the bottom of the pattern (Figure 3B), which cannot be replicated by any combination of the pattern center components. Similarly, the sensitivity field for the azimuthal angle contains some vectors with a vertical component towards the sides of the pattern—a sensitivity distinct from that of *pcx*. Note that these sensitivity vectors and couplings are unique for the initial detector parameters used here; however, similar trends and couplings are expected regardless of the initial parameters, since each parameter has a consistent displacement signature. For example, if the initial *pcx* and *pcy* were shifted right and down, the displacement field due to perturbation of *pcz* would still show a concentric dilation but centered about the shifted pattern center location. The algorithm recomputes the Jacobian matrix for each updated (current) detector geometry in the refinement process.

While all parameters can be toggled on/off for refinement, we recommend turning on at least the pattern center components and one tilt component. Deviations in the sample tilt are certainly possible due to mounting errors, and although the detector angles can usually be considered constant, there is the possibility of installation error or hardware positional shifts with usage (potentially a few tenths of a degree). Turning on all parameters may result in non-uniqueness of the solution, since several parameters are coupled (further discussed in Section 4.2). However, if the optimized geometry results in correct indexing, it does not matter if it is the physically "correct" solution.

### 2.4. Global geometry corrections

We now have the measured residual field vector $\mathbf{b}(\mathbf{p}_0)$ for the current geometry (describing the global misalignment) and the Jacobian sensitivity matrix $\mathbf{J}$ (describing the effect of perturbing each parameter). We can approximate the true detector geometry by solving for the linear combination of parameter changes that will best eliminate the global pattern misalignment. Assuming small parameter variations, this zero-residual condition can be conceptually written as

$$\mathbf{b}(\mathbf{p}_0 + \Delta \mathbf{p}) \approx \mathbf{b}(\mathbf{p}_0) + \mathbf{J}\Delta \mathbf{p} \approx \mathbf{0} \tag{6}$$

To drive the resultant residual vector $\mathbf{b}(\mathbf{p}_0 + \Delta \mathbf{p})$ towards zero in one step, we solve the general $\ell_2$-norm least-squares problem

$$\min_{\Delta \mathbf{p}} \|\mathbf{b}(\mathbf{p}_0) + \mathbf{J}\Delta \mathbf{p}\|_2^2. \tag{7}$$

Instead of solving this optimization problem as is, we introduce weights, normalization, and damping for stability and robustness. Weights are applied to the rows of $\mathbf{J}$ and $\mathbf{b}$ to account for ROIs with limited feature pairs, which could reduce the reliability of the calculated displacement vectors. The weights are computed directly from the number of vectors used in the average displacement vector calculation for that ROI ($c_r = \sum_i c_{r,i}$):

$$w_r = \begin{cases} 0, & \text{for } c_r = 0 \\ \sqrt{c_r / c_{90}}, & \text{for } 0 < c_r < c_{90} \\ 1, & \text{for } c_r \geq c_{90} \end{cases}, \tag{8}$$

where $c_{90}$ is the 90$^{\text{th}}$ percentile of all nonzero counts. The weight vector $\mathbf{w} = [w_1, w_1, w_2, w_2, \ldots, w_R, w_R]$ is applied to $\mathbf{J}$ and $\mathbf{b}$ to yield $\mathbf{J}_w = \mathbf{wJ}$ and $\mathbf{b}_w = \mathbf{wb}$. For stability, we apply column preconditioning by normalizing the values by the column vector magnitudes ($s_k = \|\mathbf{J}_w[:,k]\|_2$):

$$\tilde{\mathbf{J}} = \mathbf{J}_w \mathbf{S}^{-1}, \tag{9}$$

where $\mathbf{S}^{-1} = diag(s_1^{-1}, s_2^{-1}, \ldots, s_K^{-1})$. Finally, we solve the weighted, normalized system of equations, including a damping factor $\lambda$ for stability (usually taken as 10$^{-3}$):

$$(\tilde{\mathbf{J}}^T \tilde{\mathbf{J}} + \lambda \mathbf{I})\Delta \tilde{\mathbf{p}} = -\tilde{\mathbf{J}}^T \mathbf{b}_w. \tag{10}$$

After solving for the scaled parameter changes $\Delta \tilde{\mathbf{p}}$, we recover the physical geometry parameter updates which minimize the pattern shifts by undoing the column normalization:

$$\Delta \mathbf{p} = \mathbf{S}^{-1} \Delta \tilde{\mathbf{p}}, \tag{11}$$

since $\mathbf{J}_w \Delta \mathbf{p} = \mathbf{J}_w \mathbf{S}^{-1} \Delta \tilde{\mathbf{p}} = \tilde{\mathbf{J}} \Delta \tilde{\mathbf{p}}$. The parameter updates are used to define a new detector with new parameters ($\mathbf{p}_1 = \mathbf{p}_0 + \Delta \mathbf{p}$). Optionally, the pattern center values for each point can be updated based on projections of the newly defined geometry using *kikuchipy*'s *extrapolate_pc* function [28]. After the new detector has been defined, *kikuchipy*'s *refine_orientation* function is used to individually refine the orientations in the dataset, since the detector update changes the pattern simulation and there may be an orientation for which the simulated and experimental pattern match better (higher normalized cross-correlation (NCC)). After orientation refinement, the average NCC after refinement is compared to the average NCC before geometry refinement. If the NCC improved by more than a threshold value (typically 0.005), the process continues. The two-step refinement process (first, the global geometry update and, second, the local orientation refinement) is iteratively repeated until the NCC score converges to a maximum value (defined as an incremental increase of less than 0.005). A schematic summary of the entire algorithm is presented in Figure 4. Our method effectively decouples the global pattern misalignments (from an incorrect sample-detector geometry) from the local misalignments due to errors in orientation. After the geometry refinement is completed on a subset of *N* points in the map, the entire EBSD map can be reindexed using the optimized sample-detector geometry.

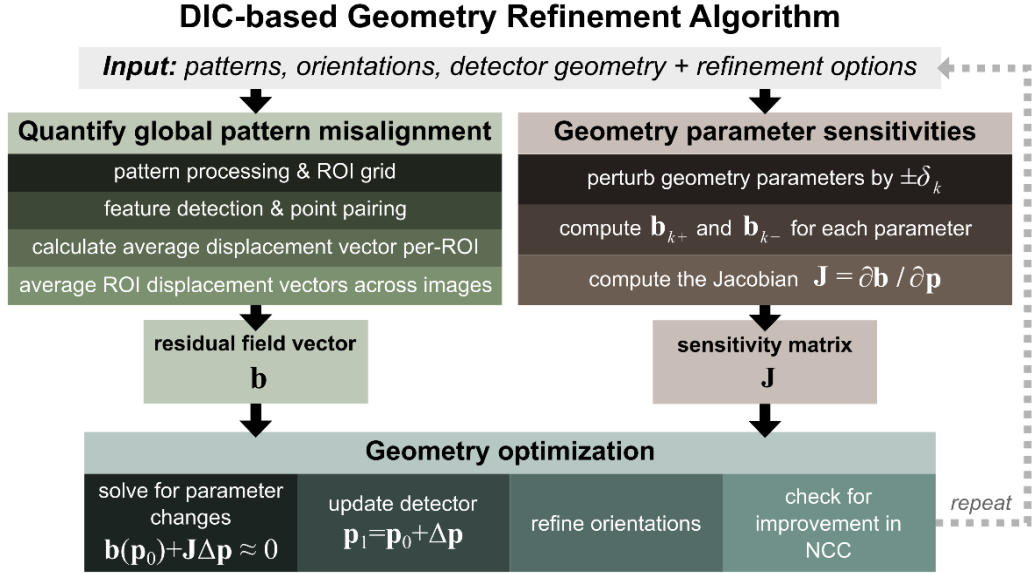

Figure 4: Schematic summary of the DIC-based geometry refinement algorithm

## 3. Results

### 3.1. Application to a silicon single crystal

We first test our geometry refinement method with a single-crystal silicon sample. A ~5x5 mm$^2$ sample was cut from a <100>-oriented, polished Si wafer (MicroChemicals GmbH). EBSD was conducted using a TFS Quanta 200F SEM equipped with an EDAX Hikari camera. Two EBSD maps were taken at ~70° sample tilt with a beam energy of 25 kV. The two maps have the same center point but different areas: the map area A2 with dimensions 10x that of map area A1 (see Figure 5A for a schematic of the map areas). The purpose of mapping one large and one small area is to test the ability of the geometry refinement algorithm with differently sized map areas. Due to beam deflection during scanning, the pattern center changes with the map geometry. This effect is more exaggerated with larger maps. In our geometry refinement method, the pattern center distribution can be updated when the sample-detector angular parameters are changed (using *Kikuchipy*'s *extrapolate_pc* function). Regardless of this spatial variation in pattern center, the two maps are expected to have the same average pattern center (pattern center at the map center) and detector and sample tilts, since the stage and detector positions were not moved between the scans (only the magnification). An indication of the two maps having the same geometry is the similarity in the initial average displacement fields (Figure 5 B and C). Although there are some differences in the magnitudes of the displacement vectors, the directions are consistent when comparing each ROI between the two datasets. This shows that the displacement fields captured by the DIC-based algorithm show consistent misalignments between the patterns and maps.

In addition to performing geometry refinement on the maps using our method, we perform pattern center/orientation refinement using local Nelder-Mead (NM) optimization and global Differential Evolution (DE) optimization algorithms from *SciPy* available within *Kikuchipy*'s *refine_orientation_projection_center* function. For these numerical optimization runs, we use a trust region of 3° for orientations and 5% detector width for pattern centers. The parameters for DE optimization are set according to the recommendations in [27]. The same points are used in geometry refinement for the three methods, taken from an equally spaced grid of 100 total points within each map. On average, it took 2.9 minutes to complete the DIC-based geometry refinement, 5.9 minutes to complete NM optimization, and 146.1 minutes to complete DE optimization using one CPU. For this example, our geometry refinement method is ~50x faster than the DE optimization method.

To assess the accuracy and robustness of the refinement strategies, we calculate the misorientation angle from the map-average orientation, which we term the *disorientation* angle. Since the sample is a single crystal, we expect all points to have nearly the same indexed orientation. Figure 5D shows colormaps of the disorientation angle for each dataset of the A1 map. Compared to the initial Hough data and the numerical optimization methods, our geometry refinement method yields the most consistent single-crystal orientations, as shown by nearly zero disorientation angles throughout the map (white map compared to red, i.e., higher disorientation angles in the other maps). The box plots in Figure 5 E and F summarize the results of all methods for both maps. Outliers are defined as any value outside of 1.5x the interquartile range (IQR). The blue box plots (left axis) show the statistics of the confidence index (CI) values: for the Hough dataset, the CI is the typical vote-based ranking metric output by the EDAX software, whereas for the refined datasets, it is the normalized cross-correlation. The CI for the Hough data is high for both areas (averages of 0.77 for A1 and 0.71 for A2), demonstrating successful identification of the correct orientation. The mean disorientation angles for the Hough datasets are 0.28° for A1 and 0.40° for A2, which are lower than the ~0.5° angular precision for conventional Hough-based indexing. However, as utilized for HR-EBSD (which requires finer angular precision), the NM and DE optimization methods decrease the average disorientation angles for both maps (0.13° for A1 and 0.17° for A2, using DE optimization). Our DIC-based geometry refinement method outperforms both numerical optimization strategies, with average disorientation angles of 0.03° for A1 and 0.14° for A2.

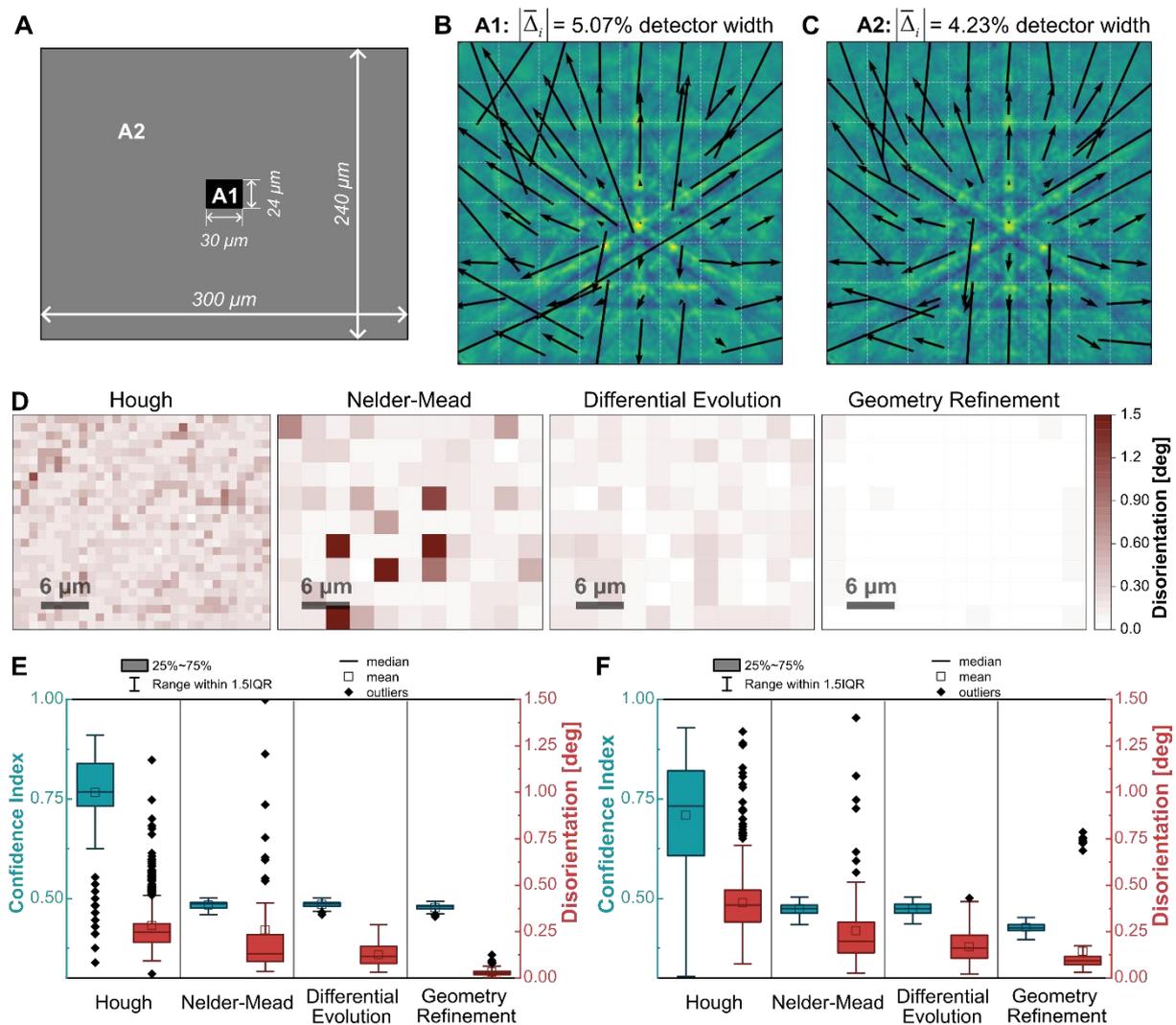

**Figure 5: Comparison of geometry refinement methods applied to a Si single crystal:** (A) Schematic showing scan areas A1 and A2 with two different sizes; (B-C) average displacement fields calculated from a sampling of 100 patterns from A1 (B) and A2 (C); (D) pixelwise maps of the disorientation (misorientation angle from map-average orientation) for the initial Hough data and the geometry-refined maps; (E-F) statistics of the confidence index and disorientation values for the initial Hough data and the geometry-refined maps for A1 (E) and A2 (F).

After demonstrating the success of our algorithm in recovering orientations consistent with a single crystal, let us compare the values of the refined geometry parameters across methods. Table 1 and Table 2 list the geometry parameters from each method for maps A1 and A2, respectively. Table 1 shows that the average *pcx* values are consistent across refinement strategies, shifting ~1.5% of the detector width from the initial *pcx* value. The average *pcy* and *pcz* values from the NM and DE optimizations are the same, while the values from our geometry refinement method are lower. However, the sample and detector tilts also changed, which are highly coupled to *pcy* and *pcz* (Figure 3C). For A2, the refined pattern center values from the NM and DE optimizations are consistent with the values from A1, revealing that the same geometry was recovered. The values from our geometry refinement method differ between the two maps. However, due to coupling between several of the parameters, the different solutions may represent two equally effective descriptions of the sample-detector geometry. In Section 4.2, we further discuss how coupling may lead to non-unique solutions.

| Method | pcx | pcy | pcz | $\theta_d$ | $\theta_s$ | $\alpha_d$ |
|---|---|---|---|---|---|---|
| Initial | 0.505 | 0.205 | 0.81 | | | |
| Nelder-Mead | 0.52 | 0.176 | 0.858 | 10.0 | 70.1 | -2.0 |
| Differential Evolution | 0.521 | 0.176 | 0.858 | | | |
| Geometry Refinement | 0.52 | 0.172 | 0.849 | 10.366 | 69.897 | -1.401 |

Table 1: Detector geometry parameters for A1, using different refinement methods

| Method | pcx | pcy | pcz | $\theta_d$ | $\theta_s$ | $\alpha_d$ |
|---|---|---|---|---|---|---|
| Initial | 0.505 | 0.205 | 0.81 | | | |
| Nelder-Mead | 0.519 | 0.176 | 0.858 | 10.0 | 70.1 | -2.0 |
| Differential Evolution | 0.52 | 0.176 | 0.858 | | | |
| Geometry Refinement | 0.507 | 0.127 | 0.835 | 10.674 | 74.222 | -1.518 |

Table 2: Detector geometry parameters for A2, using different refinement methods

To further analyze the pattern center optimization between the methods, we plot surface maps of *pcx*, *pcy*, and *pcz* which show the optimized pattern center for each point in the map (Figure 6). Flat planes represent the pattern center distributions for our geometry refinement method, since refinement is performed globally and the spatial variation of pattern centers is calculated from the geometry. In contrast, the point-wise numerical optimization methods show large point-by-point deviations in the pattern centers. These spatial variations reflect the speckled nature of the disorientation plots (Figure 5D) and demonstrate the non-uniqueness due to coupling during simultaneous orientation and pattern center refinement. Ideally, these surface plots should be close to planar like those for our geometry refinement method. For the larger map, the pattern center distributions for the numerical optimization methods seem to follow the correct geometry-prescribed distributions in an average sense. However, there are still large local deviations which translate to errors in the orientations.

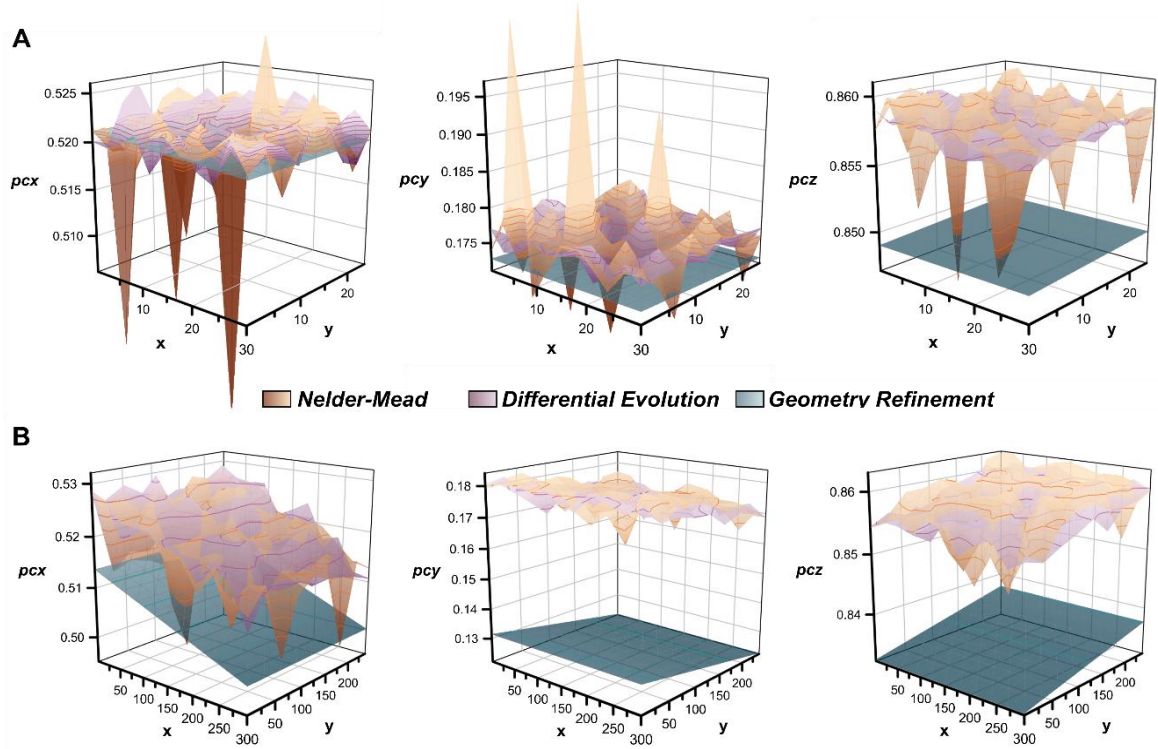

**Figure 6: Calibrated pattern center maps:** Refined *pcx*, *pcy*, and *pcz* values as a function of position for each geometry refinement strategy for map A1 (A) and A2 (B).

*3.2. Application to a material with pseudo-symmetry*

While single-crystal silicon is a useful test case, it does not demonstrate the potential of our method to overcome current challenges in EBSD analysis where correct sample-detector geometry is essential. Here, we apply our method to a single-crystal barium titanate sample (BTO)—a ferroelectric material possessing microstructural domains of different polarization directions as defined by pseudosymmetric relations of the tetragonal unit cell. For a tetragonal ferroelectric, there are six possible polarization states within the unit cell. To accurately distinguish between the Kikuchi patterns of these variants, we developed a pseudo-symmetry sensitive EBSD reindexing technique [31]. The technique relies on orientation refinement using simulated pattern-matching and a novel confidence index that considers not only pattern similarity but also pattern dissimilarity trends between PS variants. Since the Kikuchi patterns of 180° domain variants are extremely similar (perfect simulated patterns have NCC scores up to 0.995), this PS-sensitive confidence index is necessary to distinguish between the patterns. However, to utilize this PS-sensitive confidence index, the Kikuchi bands of the experimental and simulated patterns must be very well aligned: meaning that accurate indexing is reliant on precise determination of the sample-detector geometry.

We test our geometry refinement method, as well as the two common numerical optimization methods, on an EBSD map of a single-crystal BTO sample. Approximately 100 points from the map were selected for geometry refinement. Manual selection of points on a map of the PRIAS signal, which shows the domain contrast, ensures good representation from each domain type so that one variant does not bias the refinement (Figure 7A). The displacement field signature from the first step of the DIC-based geometry refinement is shown in Figure 7B. The average displacement field shows a coherent pattern, reflecting a consistent misalignment between experimental and simulated patterns due to a sample-detector geometry issue. The results of the three geometry refinement methods are shown through the statistical box plots of the NCC and disorientation angle in Figure 7C. Since up to six PS orientation variants may be represented in the datasets, we compute the disorientation angle using reduced cubic symmetry. The box plots show that the DE optimization outperforms the NM optimization with lower disorientation angles and higher NCC values, and our geometry refinement algorithm outperforms both with the lowest average disorientation angles and highest NCC values. The disorientation angles are higher than that for the silicon sample, but this is expected since six orientation variants are represented and there may be small lattice rotation between domains.

| Method | pcx | pcy | pcz | $\theta_d$ | $\theta_s$ | $\alpha_d$ |
|---|---|---|---|---|---|---|
| **Initial** | 0.501 | 0.202 | 0.893 | | | |
| **Nelder-Mead** | 0.513 | 0.156 | 0.882 | 10.0 | 69.9 | -2.0 |
| **Differential Evolution** | 0.513 | 0.153 | 0.881 | | | |
| **Geometry Refinement** | 0.503 | 0.125 | 0.867 | 9.382 | 73.129 | -1.851 |

Table 3: Detector geometry parameters for BTO map using different refinement methods

The parameters for the initial geometry, NM and DE optimized geometries, and the geometry from our DIC-based refinement method are shown in Table 3. The sample and detector angles were modified with our method, along with the pattern center values. The parameter values shown in the table were used to create new detectors for the full map with spatially varying pattern centers populated using *Kikuchipy*'s *extrapolate_pc* function. These three detectors were then used with our PS-sensitive EBSD reindexing method [31] to solve for the orientations and distinguish between variants.

Figure 7D shows the results of the PS-sensitive reindexing using each refined detector. The top colormaps show the domain variant of each point, using the color-code shown on the right-hand side. The maps refined with the NM and DE optimized detectors are noticeably noisier, with the domains not having a consistent variant designation than the map refined with our geometry-refined detector. The grayscale maps below show the cross-variant margin (CVM), which is the difference in the confidence index values between the two top variants. The maps refined with the NM and DE optimized detectors have lower CVM values throughout, revealing that the discrimination between the variants is poorer. Finally, in [31] we confirm the domain variant orientations are correctly identified using our EBSD reindexing technique by comparing the results to piezoresponse force microscopy scans taken of the same area. Thus, we can state that our geometry refinement algorithm identifies a geometry that is globally consistent with the dataset and enables accurate discrimination between close PS variants. The geometries refined using numerical optimization methods produce erroneous results, including the identification of incorrect domain variants (for example, the light green domains identified as dark red).

In part, this can stem from the point-wise optimization, which does not decouple the local orientation (or variant) changes from the global geometry.

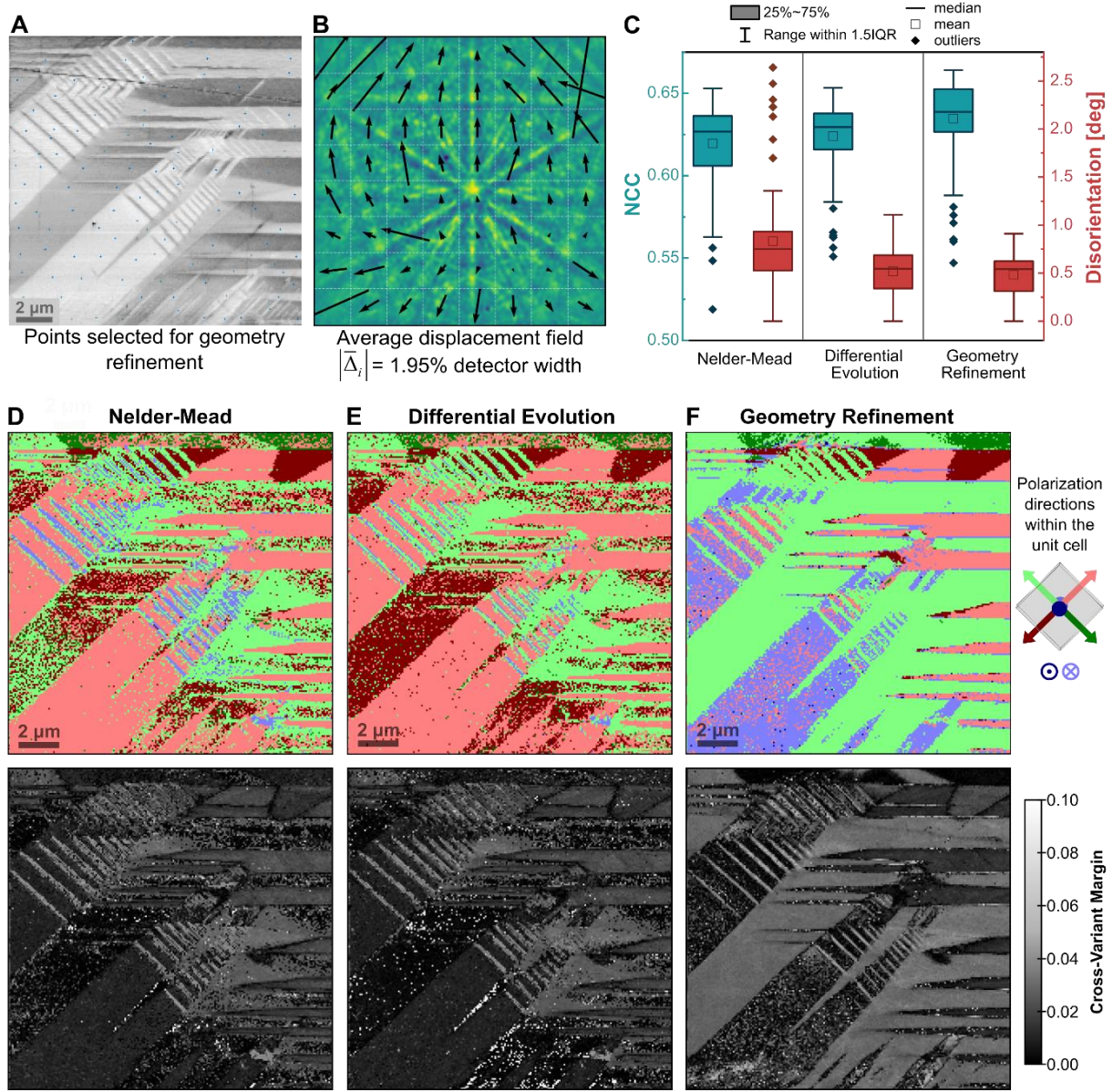

**Figure 7: Geometry optimization for accurate PS variant discrimination in BTO:** (A) PRIAS signal of EBSD map with points selected for geometry refinement identified by blue crosses; (B) average displacement field calculated during the first DIC-geometry refinement step; (C) statistics of the normalized cross-correlation and disorientation angle for each geometry refinement method; (D) polarization direction maps (top) and cross-variant margin (bottom) for each refinement method.

## 4. Discussion

### 4.1. Validity range of linear approximations

Although our geometry refinement algorithm assumes a linear combination of parameter sensitivities can account for the pattern shifts, the relationship is highly nonlinear since the patterns are formed from projections of 3D diffraction cones, and the geometry parameters have coupled effects. To test the validity range of the linear approximation used for geometry refinement, we create synthetically perturbed datasets of simulated patterns. To generate a synthetic "base" dataset ($\mathbf{A}^i$), we use the BTO master pattern, an initial geometry ($\mathbf{p}_i$), and the ~100 orientations used for geometry refinement of the BTO single crystal dataset in Figure 7A. (In principle, any material, orientations, and initial detector could be used for this analysis.) We perturb the geometry by a prescribed $\Delta\mathbf{p}$ and simulate a new dataset of patterns with the same orientations but using detector parameters $\mathbf{p}_t = \mathbf{p}_i + \Delta\mathbf{p}$. This dataset becomes a synthetic "experimental" dataset (**B**) whose geometry is different from the initial (known) sample-detector geometry parameters. We then run the geometry refinement algorithm to solve for geometry parameters ($\mathbf{p}_r$) that bring the "base" patterns in alignment with the "experimental" patterns, yielding a new simulated dataset of patterns ($\mathbf{A}^r$).

With six geometry parameters, a six-dimensional space must be analyzed to find the bounds of validity for the linear approximation. For each parameter, we select five values (0, $\pm dp/2$, $\pm dp$) where $dp = \{0.05, 0.05, 0.08, 1, 1, 3\}$ (for the parameters in the order of $\mathbf{p}$). The first three $dp$ values are deviations in the pattern center values, which could reasonably be up to 5% of the detector width. We assume maximum deviations of 1° in the detector angles and 3° in the sample tilt. With five values for six parameters, the number of combinations is $5^6$. Therefore, to reduce the number of combinations we only test the pairwise combinations (or 375 cases). For each of the 15 parameter pairs (sub-diagonal of the parameter coupling matrix in Figure 8B), 25 cases of perturbed geometries are run with $\Delta p$ values only for the two selected parameters, while all other $\Delta p$ values remain zero.

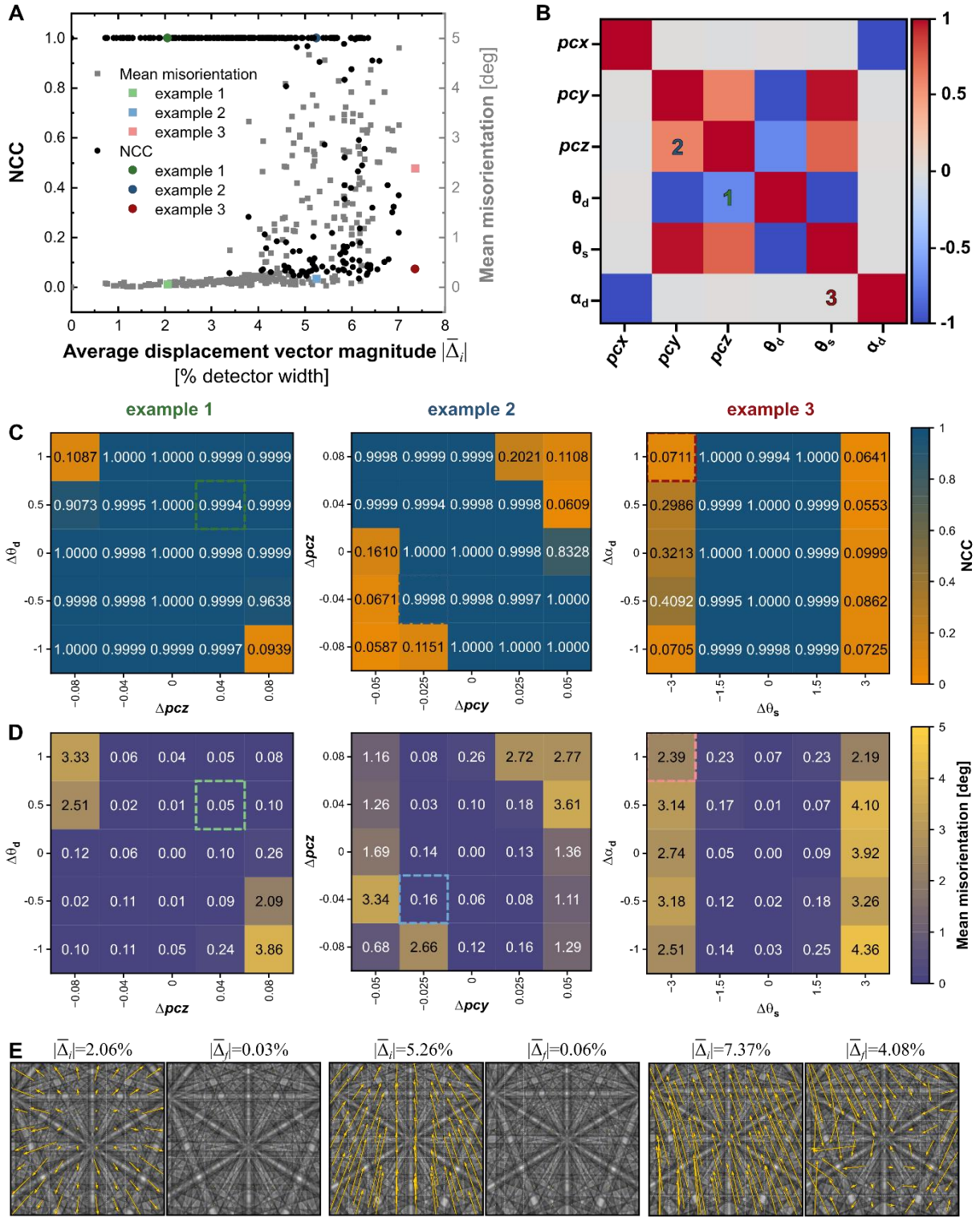

Figure 8: Validity bounds of parameter perturbations: (A) Average NCC after geometry refinement and average misorientation of refined dataset to true orientations as a function of the initial average displacement vector magnitude (data for three examples are highlighted); (B) coupling matrix with the parameter pairs for the three examples identified in (A); (C) NCC after geometry optimization as a function of two parameter perturbations (selected examples identified with dashed borders); (D) mean misorientation (to true orientations) as a function of two parameter perturbations (selected examples identified with dashed borders); (E) average displacement fields before (left) and after (right) geometry refinement with the total average displacement magnitudes labeled above images.

We use two metrics to assess the success of the geometry refinement: (i) the average normalized cross-correlation (NCC) between the refined dataset $\mathbf{A}^r$ and the "experimental" dataset $\mathbf{B}$, and (ii) the mean misorientation between the true (initial) orientations and the refined orientations. For geometry refinement to be successful, the NCC should be nearly 1 and the mean misorientation should be nearly 0. In an attempt to collapse the 6D geometry parameter perturbation space into one useful metric, we compute the average displacement magnitude from the displacement field describing the shifts between the initial pattern sets ($\mathbf{A}^0$ and $\mathbf{B}$). The average displacement vector magnitude is normalized by the detector width. Figure 8A shows the average NCC and misorientation values as a function of the average displacement magnitude for all 375 cases tested. Below an average displacement magnitude of about 4% of the detector width, the NCC is consistently close to 1 and the misorientation is consistently close to 0—demonstrating the success of the geometry refinement algorithm at recovering the correct geometry while preserving the correct orientations. This further demonstrates that the effects of global geometry are effectively decoupled from the local orientation effects. Above an average displacement magnitude of 4%, there are some cases with low NCC and high misorientation (up to 5 degrees). In these cases, the geometry refinement fails. However, there are some cases with initial displacement magnitudes above 4% for which the geometry refinement is still successful. Given that numerical optimization methods typically operate in ±1-5% detector-width search basins for the pattern center optimization [27], a 4% window of reliability is sufficient for most EBSD datasets. Unless the EBSD camera calibration is highly erroneous, the initial detector geometry parameters read from the EBSD system should not give such large pattern displacements.

In Figure 8A, we have also highlighted the NCC and misorientation values for three example cases. The parameter pairings used for these three example cases are shown in the parameter coupling matrix of Figure 8B. Heat maps showing the average NCC and misorientation values for each of the 25 perturbation combinations for each example parameter pairing are shown in Figure 8C and Figure 8D, respectively. The initial and final (after geometry refinement) average displacement fields for each example are shown in Figure 8E. Example 1 is a case where the detector tilt and *pcz* are perturbed together, which are inversely coupled parameters (Figure 8B). The NCC and misorientation heat maps show that most perturbation combinations of these two parameters result in successful geometry refinement (with high NCC values and low misorientations), except for the upper left and bottom right corners. These regions correspond to two extremes of the parameter combinations—where $\Delta pcz$ is minimum (-0.08) and $\Delta \theta_d$ is maximum (1) and vice versa. Since these two parameters are inversely coupled, applying a positive perturbation of one and a negative of the other produces an additive effect on the pattern displacement field. For the specific example case, $\Delta pcz = 0.04$ and $\Delta \theta_d = 0.5$ creates the average displacement field in Figure 8E (left) with an average displacement magnitude of 2.06%. The geometry refinement for this case is successful, with a final average displacement magnitude of 0.03%, an average NCC of 0.9994, and an average misorientation of 0.05 degrees. Since the initial average displacement magnitude is less than 4%, the geometry refinement method works predictably well with this case.

Example 2 provides a case where the geometry refinement still works excellently (NCC=0.9998, misorientation=0.16 degrees) despite an initial displacement magnitude above 4% (Figure 8E middle, 5.26%). The two perturbed parameters for this case, *pcy* and *pcz*, are positively coupled. Since negative perturbations are applied to both ($\Delta pcz = -0.04$ and $\Delta pcy = -0.025$), their effects are additive, which explains the high initial displacement magnitude. Nevertheless, the algorithm still performs well in this case. At further extremes (high/low $\Delta pcz$ and $\Delta pcy$), the algorithm clearly fails, as indicated by low NCC and high misorientation values. This again reflects the coupling between the parameters, where in this case, perturbations of the same sign result in additive pattern displacement effects.

Example 3 demonstrates a scenario without parameter coupling. The two perturbed parameters are the sample tilt (which mostly produces a *y*-directional displacement field) and the detector azimuthal angle (mostly producing an *x*-directional displacement field). Unlike the other two examples with parameter coupling, the refinement does not fail at extreme combinations of both parameters but rather at the extremes of the sample tilt. This demonstrates that the selected extremes of the azimuthal angle (±1 degree) were reasonable, but the selected extremes of the sample tilt were too high (±3 degrees). We examine a case ($\Delta \theta_s = -3$ and $\Delta \alpha_d = 1$) where the initial displacement is high (7.37%) and the geometry refinement fails. The failure of the geometry refinement algorithm in this case is evident by the low NCC value (0.0711), relatively high misorientation (2.39 degrees), high displacement magnitude (4.08%), and incoherent displacement field after refinement (Figure 8E, right). This failure demonstrates that the large pattern shifts caused by perturbing these uncoupled parameters are likely highly nonlinear and cannot be reconciled by a linear combination of parameter sensitivities.

### 4.2. Parameter error and non-unique solutions due to parameter coupling

In the previous analysis, we synthetically perturbed only two parameters at a time but then let all parameters evolve during geometry refinement. This scenario reflects reality, in which all parameters must be allowed to evolve, since it is generally not known which parameters are incorrect and must be calibrated. Although necessary, allowing all parameters to evolve can lead to non-unique solutions, since there is strong coupling between some parameters. For example, since *pcy* and the sample tilt are strongly coupled, a positive perturbation in one may be compensated for by a negative perturbation in the other parameter. This potential for several non-unique solutions does not necessarily result in incorrect geometry calibration. Rather, there may be several ways to describe the correct geometry.

To assess the effects of parameter coupling on geometry calibration, we perform error analysis on the calibrated geometries for the synthetically perturbed datasets from Section 4.1. The "true" geometry ($\mathbf{p}_t$) is compared to the refined geometry ($\mathbf{p}_r$) to calculate the error in each parameter ($\varepsilon_k = p_k^r - p_k^t$). To compare errors across parameters, we calculate the normalized error:

$$\hat{\varepsilon}_k = \varepsilon_k \frac{\bar{d}_N}{2\delta_k}, \tag{12}$$

where $\bar{d}_N$ is the average displacement vector magnitude normalized by the detector width, which is calculated from the average displacement field (Figure 3B) obtained by perturbing geometry parameter $k$ by $\pm\delta_k$. With this normalization strategy, we can directly compare the errors between parameters. In Figure 9A and B, we plot the normalized parameter errors for Examples 1 and 3 presented in Section 4.1. The stacked columns represent the total error for each of the 25 perturbation combinations. On the "floor" of the 3D column charts are the NCC heatmaps presented in Figure 8C. For both examples, the total error is significantly larger for the cases where the NCC is low, denoting a failed calibration. The total error in these failed cases tends to be dominated by errors in the "active" parameters (errors in *pcz* and $\theta_d$ are large for the cases with low NCC in Figure 9A; errors in $\theta_s$ and $\alpha_d$ are large for the cases with low NCC in Figure 9B). Large errors in the active parameters for the failed calibration cases show that the perturbed geometry was not properly identified and corrected during geometry refinement. In these failed cases, there are also appreciable errors in some of the other unperturbed parameters, illustrating that these parameters were modified during geometry refinement even though they should have been kept constant.

While it is expected that large errors in the geometry parameters are associated with unsuccessful refinement cases (low NCC score), some of the successful refinement cases shown in Figure 9A (NCC≈1) have appreciable errors. The parameters which have the most error in these cases are the detector tilt (purple) and sample tilt (blue). In each case, the normalized error for each of these parameters is roughly equal and of the same sign. Since these two parameters are strongly inversely coupled, the errors effectively negate each other. This demonstrates the hypothesized potential for non-unique solutions: although the optimized values of these two parameters are not what was prescribed, the strong coupling allows for alternative sets of parameter values which effectively describe the sample-detector geometry.

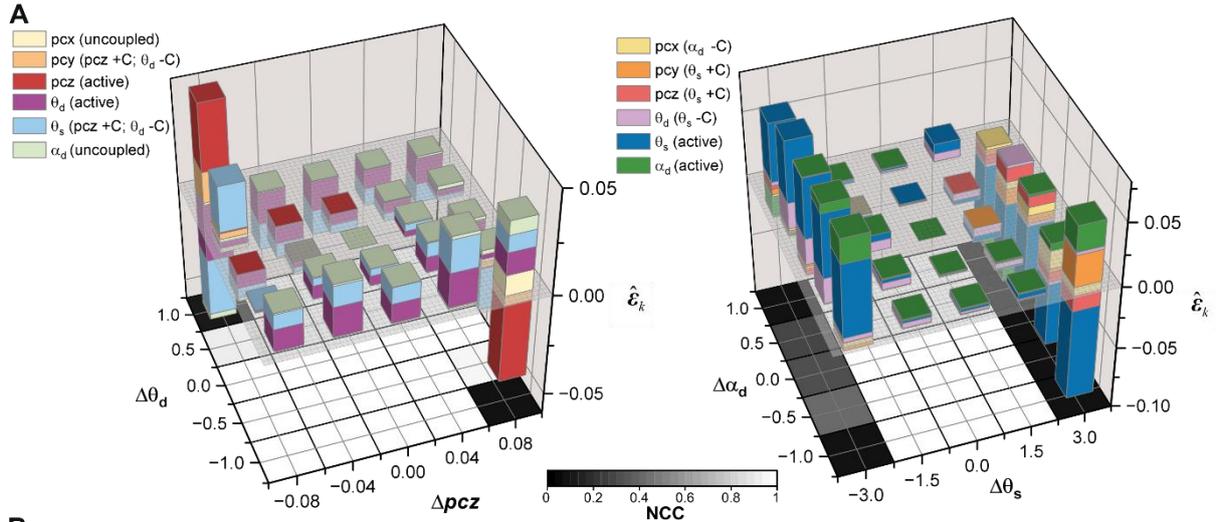

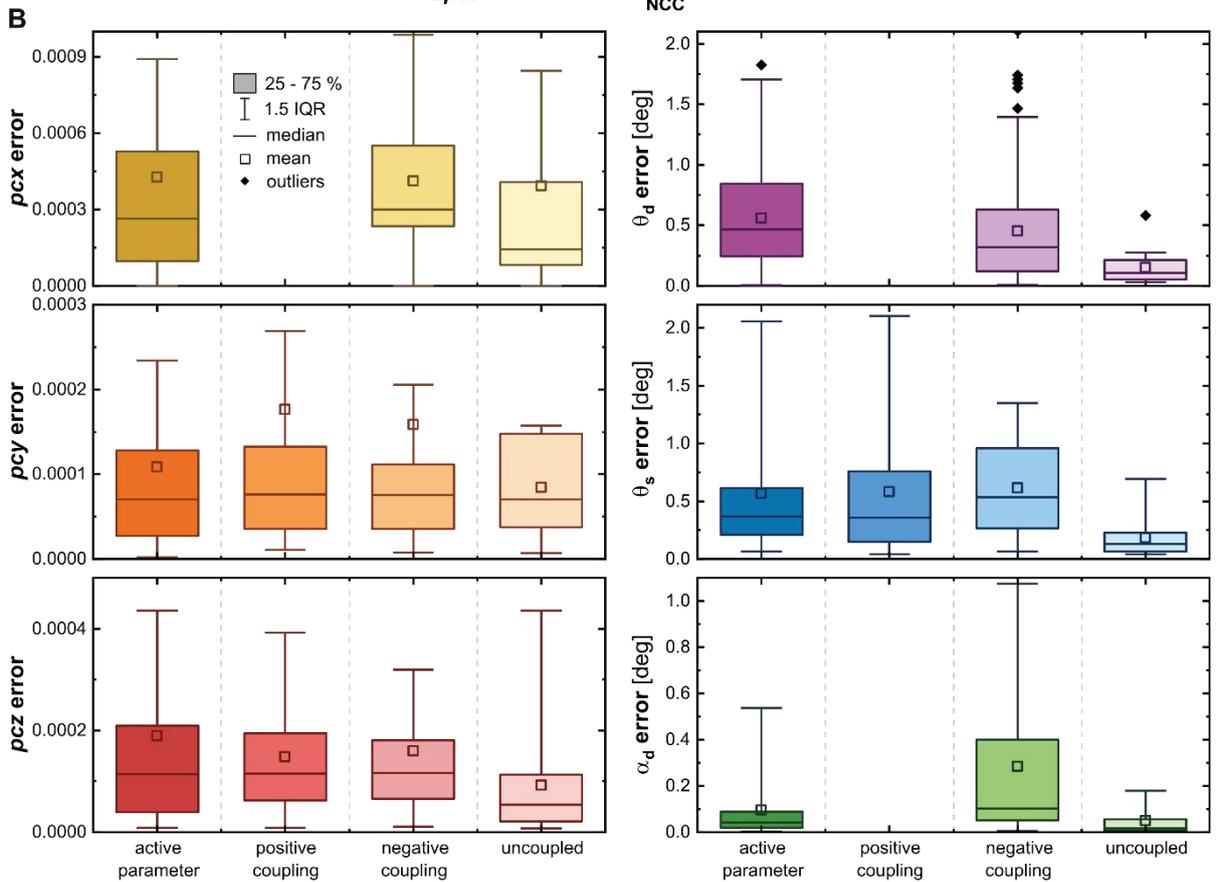

**Figure 9: Geometry parameter errors:** (A) Normalized error of each geometry parameter as stacked columns for the 25 geometry refinement runs for example 1 (left: *pcz* and detector tilt) and example 3 (right: sample tilt and azimuthal), with the NCC value for each refinement run indicated by the grayscale floor tile; (B) statistics for the absolute error in each parameter, grouped the parameter's status in the refinement run.

In Figure 9B, we report the statistics for the absolute error $|\varepsilon_k|$ in each parameter. Only successful refinement cases (defined as having a final NCC > 0.9) from Section 4.1 are included (258 out of 375 total cases). The calculated parameter errors for all cases are consolidated and grouped by the parameter's status in the refinement run. If the parameter was one of the two perturbed in the run, it is marked as an "active parameter". If the parameter is coupled positively (red in coupling matrix of Figure

3C) or negatively (blue in coupling matrix of Figure 3C) to one of the active parameters in the run, it is grouped accordingly. The final category "uncoupled" contains the errors of parameters which have no coupling with either active parameter. The box plots in Figure 9B show the statistics for each of these parameter groupings. For all parameters, the "uncoupled" groups have the lowest median and mean values compared to the active or coupling groups for that parameter, revealing that the refinement algorithm is generally correctly neglecting these uncoupled parameters which have limited contribution towards correcting the global pattern misalignment (residual field vector). For the coupling groups (negative and positive), the mean and median errors are comparable to, and in some cases higher than, the respective values for the active parameter group. This again demonstrates that parameter coupling results in non-unique solutions for the refined geometry.

Finally, we analyze the values of the parameter errors in comparison to those reported for state-of-the-art geometry refinement strategies. The pattern center errors are on the order of $10^{-4}$ (or 0.01% detector width). This is on-par with the 0.01% mean accuracy in pattern center reported in [27] when using DE optimization. Peng et al. reported errors as low as 0.0002–0.008%, using their global optimization algorithm [26], This, however, relies on averaging, which is not possible for large maps or materials with close pseudosymmetries. Additionally, our method also refines the sample and detector angles, so changes in these parameters may account for some of the error in the pattern center due to parameter coupling.

## 5. Conclusions

In this work, we have presented a new method for EBSD geometry calibration. Unlike most calibration techniques, our method refines not only the pattern center components but also the detector and sample angles. The foundation of the method lies in uncoupling the local crystallographic changes from the global sample-detector geometry. We accomplish this by performing DIC on simulated-experimental pattern pairs to identify the consistent map-wide pattern shifts that are associated with global geometry errors. Linear decomposition of this geometry-error signature into contributions from each parameter allows us to recover a globally consistent effective geometry for the dataset. We have demonstrated the success of this method on single-crystal silicon and barium titanate samples, achieving an average misorientation from the mean orientation of 0.03° in silicon and correctly identifying the pseudosymmetric domain variants in barium titanate. We demonstrate the improved accuracy of our method over typical numerical optimization strategies, with the added benefit of performing up to 50x faster. In contrast to independent pattern-wise optimization strategies, the proposed method reduces the coupling between local orientations and the global geometry, as evidenced by the lower disorientation angles and smoother geometry fields. Additionally, we show that our method can correct for errors in the sample and detector angles, which would normally be compensated for by erroneous offsets in the crystal orientations. While incorporating these geometry angles in refinement may improve orientation accuracy, these parameters are highly coupled to the pattern center components and may result in non-unique geometry solutions. Nevertheless, we show that pattern center errors on the order of 0.01% in detector width can be achieved despite this geometry parameter coupling. While proven to be a promising new method for EBSD geometry calibration, there is room for improvement. First, the linear approximation is shown to break down when the average pattern displacement magnitude exceeds 4% of the pattern width. A nonlinear model of geometry parameter effects could extend the applicability of the method. Furthermore, improvements can be made to the DIC approach to be more robust to pattern noise. We encourage adoption and innovation of this promising global geometry refinement technique within the EBSD community.

**Data and Code Availability**

Upon publication, all EBSD data will be made publicly available, hosted on ETH Zurich's Research Collection. The developed code will be made publicly available on GitHub.

**Acknowledgements**

The authors gratefully acknowledge the helpful discussions and support from Mathieu Brodmann, Hsu-Cheng Cheng, Christian Franck, and Vignesh Kannan. The authors gratefully acknowledge ScopeM for their support and assistance in this work. The authors gratefully acknowledge the financial support from the Swiss National Science Foundation (SNSF) under project 212643.

**Declaration of generative AI and AI-assisted technologies in the manuscript preparation process**

During the preparation of this work, the authors used ChatGPT to assist with code prototyping and debugging, and ChatGPT with Scite to assist in evaluating literature coverage during manuscript revision. After using this tool/service, the author(s) reviewed and edited the content as needed and take(s) full responsibility for the content of the published article.